\documentclass[aps,prl,reprint,superscriptaddress,showpacs,nofootinbib]{revtex4-1}

\usepackage{color}
\usepackage{footnote}
\usepackage{hyperref}
\usepackage{epsfig}
\usepackage{changes}
\usepackage{lipsum}
\definechangesauthor[name={Per cusse}, color=blue]{per}
\setremarkmarkup{(#2)}

\begin{document}


\title{High Flux Femtosecond X-ray Emission from the Electron-Hose Instability in Laser Wakefield Accelerators}


\author{C. F. Dong}
\affiliation{Department of Astrophysical Sciences and Princeton Plasma Physics Laboratory, Princeton University, Princeton, NJ 08544, USA}
\affiliation{Center for Ultrafast Optical Science, University of Michigan, Ann Arbor, MI 48109, USA}

\author{T. Z. Zhao}
\affiliation{Center for Ultrafast Optical Science, University of Michigan, Ann Arbor, MI 48109, USA}
\author{K. Behm}
\affiliation{Center for Ultrafast Optical Science, University of Michigan, Ann Arbor, MI 48109, USA}

\author{P. G. Cummings}
\affiliation{Center for Ultrafast Optical Science, University of Michigan, Ann Arbor, MI 48109, USA}

\author{J.~Nees}
\affiliation{Center for Ultrafast Optical Science, University of Michigan, Ann Arbor, MI 48109, USA}
\author{A.~Maksimchuk}
\affiliation{Center for Ultrafast Optical Science, University of Michigan, Ann Arbor, MI 48109, USA}

\author{V.~Yanovsky}
\affiliation{Center for Ultrafast Optical Science, University of Michigan, Ann Arbor, MI 48109, USA}

\author{K. Krushelnick}
\affiliation{Center for Ultrafast Optical Science, University of Michigan, Ann Arbor, MI 48109, USA}
\author{A. G. R. Thomas}
\affiliation{Center for Ultrafast Optical Science, University of Michigan, Ann Arbor, MI 48109, USA}
\affiliation{Physics Department, Lancaster University, Lancaster, United Kingdom, LA1 4YB}



\begin{abstract}
Bright and ultrashort duration X-ray pulses can be produced by through betatron oscillations of electrons during Laser Wakefield Acceleration (LWFA). Our experimental measurements using the \textsc{Hercules} laser system demonstrate a dramatic increase in X-ray flux for interaction distances beyond the depletion/dephasing lengths, where the initial electron bunch injected into the first wake bucket catches up with the laser pulse front and the laser pulse depletes. A transition from an LWFA regime to a beam-driven plasma wakefield acceleration (PWFA) regime consequently occurs. The drive electron bunch is susceptible to the electron-hose instability and rapidly develops large amplitude oscillations in its tail, which leads to greatly enhanced  X-ray radiation emission. We measure the X-ray flux as a function of acceleration length using a variable length gas cell. 3D particle-in-cell (PIC) simulations using a Monte Carlo synchrotron X-ray emission algorithm elucidate the time-dependent variations in the radiation emission processes. 
\end{abstract}

\pacs{XXX}

\maketitle

Laser Wakefield Acceleration (LWFA) is a technique that has experimentally demonstrated electron acceleration to greater than GeV energies over cm-scale acceleration lengths \cite{Wang2013,Leemans2014}. In LWFA, electrons are accelerated by strong electric fields associated with plasma waves generated by a high-power, short pulsed laser. These electrons can produce energetic X-ray emission via oscillations of the electron beam in the transverse fields of the plasma wake \cite{Corde2013} and may also be used as a diagnostic of the electron dynamics in the plasma wake \cite{Corde2012,Kneip2012,Schnell2012}. Experimental measurements of such radiation generated by betatron oscillations in an LWFA have demonstrated bright and well-collimated beams of X-rays \cite{Rousse2004,Phuoc2005,Kneip2010}. The X-ray generation mechanism has been studied theoretically in an ion channel \cite{Esarey2002,Kostyukov2003,Phuoc2008a,Phuoc2008b} or bubble \cite{Thomas2009,Thomas2010}. These X-ray sources are particularly appealing for ultrafast imaging applications due to their femtosecond duration \cite{Lundh2011}. 

Recent studies have shown that a transition from an LWFA to a beam-driven plasma wakefield accelerator (``PWFA'') occurs when the first group of electrons catches up with the laser front, following depletion of the laser pulse and dephasing \cite{Masson2015}. The transverse ponderomotive force of the laser pulse no longer sustains the wakefield, which is now generated by the electron beam. Deviations of the electron beam through interaction with the weak laser field plants the seed for the electron-hose instability \cite{Whittum1991}. In general, this instability leads to spatiotemporally growing oscillations of the beam centroid along its axis of propagation and can limit the useful acceleration length, as well as increasing the beam emittance. Later studies that investigated this instability for beam-driven plasma accelerators showed that the growth rate was reduced when non-adiabatic ion-channel formation and magnetized plasma was taken into account \cite{Dodd2002,Huang2007}. For typical LWFA experimental conditions, however, the growth rate can be sufficiently rapid to produce violent electron oscillations in the laser polarization direction that eventually cause beam break-up and termination of the wakefield \cite{Huntington2011}. Although these instabilities are undesirable for some plasma based accelerator applications, they will increase the radiated flux dramatically due to the large-amplitude transverse oscillations of the electron beam resulting from this instability. A similar phenomenon is laser beam hosing \cite{Kaluza2010}, which has been shown in simulations to increase X-ray flux \cite{Ma2016}. Another mechanism is resonant interaction of the electron beam with the laser fields, which has also been shown to enhance the X-ray production \cite{Thomas2009, Cipiccia_NP_2011}. Although the X-rays produced by the instability have a larger source size relative to those produced in the laser driven wakefield, they maintain their femtosecond duration and have much larger flux and therefore may be useful for some ultrafast pump-probe applications.


In this paper, we measure the X-ray flux as a function of acceleration length in an LWFA using a variable length gas cell with fixed density. We show that for propagation extending many times the dephasing length, where the electron beam is susceptible to the electron-hose instability \cite{Lau1989,Whittum1991,Huang2007,Huntington2011}, the X-ray flux is consequently increased dramatically due to the violent wiggling of the high energy electron beam. The propagation length-dependent variations in betatron radiation are also demonstrated in detail by 3-D \textsc{Osiris} particle-in-cell simulations with a Monte Carlo synchrotron X-ray emission algorithm.

\begin{figure}[!ht]
\centering
\includegraphics[width=0.5\textwidth]{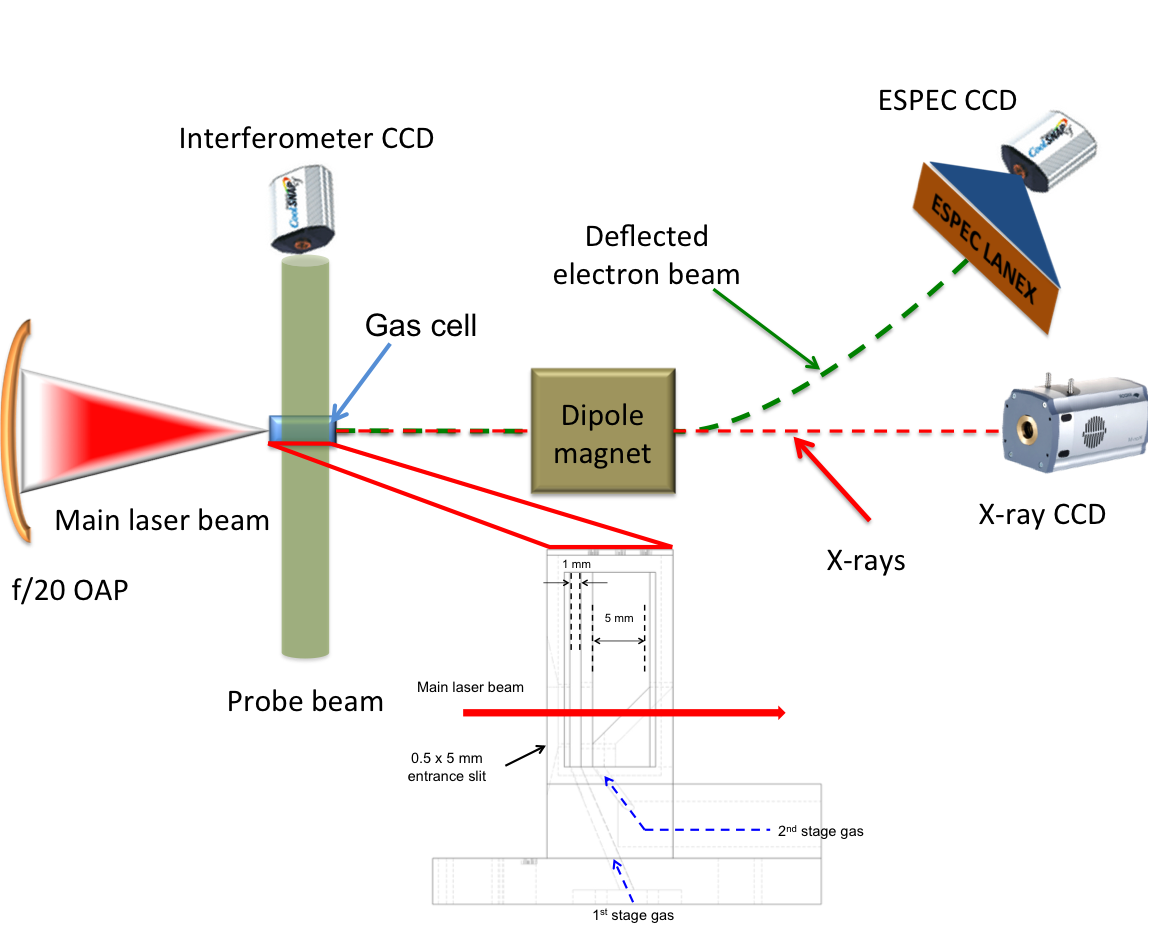}
\caption{\label{fig:setup} Schematic of the experimental setup. The main beam is focused using an $f$/20 off-axis paraboloid onto a two-stage gas cell. Electrons exiting the gas cell were swept by a dipole magnet and imaged using a LANEX screen in conjunction with a CCD camera. X-rays were measured 2.5 m downstream with an X-ray camera. Two stage, variable length gas cell is zoomed in. The acceleration length can be changed by adjusting the vertical height of the gas cell with respect to the main laser beam (as a result of the 45 degree incline in the second stage). The dashed arrows show the path of gas flow into the cell.}
\end{figure}

The experiment was performed using the 800 nm wavelength, Ti:Sapphire \textsc{Hercules} laser facility at the University of Michigan. Fig.~\ref{fig:setup} shows the experimental setup. The horizontally polarized laser beam, delivering 1.6 J in a 34 fs pulse (full-width at half-maximum, FWHM, in intensity), was focused using an $f$/20 off-axis paraboloidal mirror to a vacuum beam waist $w_0$ = 26 $\mu$m (at 1/e$^2$ of peak intensity) with an on-target peak intensities  of 2 $\times$ 10$^{19}$ W cm$^{-2}$, corresponding to a normalized field strength $a_0 = 3$. A 3D-printed variable-length two-stage gas cell was used as the target \cite{Vargas2014}. The cell was composed of a 1 mm injection stage, a 0.5 mm divider slit for stage separation, and an adjustable 0--5~mm acceleration stage, which can be adjusted by varying the height of the gas cell. The injection stage was filled with a mixed gas (97.5\% He and 2.5\% N$_2$) \cite{McGuffey2010} and the acceleration stage with pure helium. In this experiment, the average peak electron density, determined interferometrically, across the entire gas cell was fixed at $(1.0\pm0.2) \times10^{19}$ cm$^{-3}$ for all lengths.

\begin{figure}[!ht]
\centering
\includegraphics[width=0.5\textwidth]{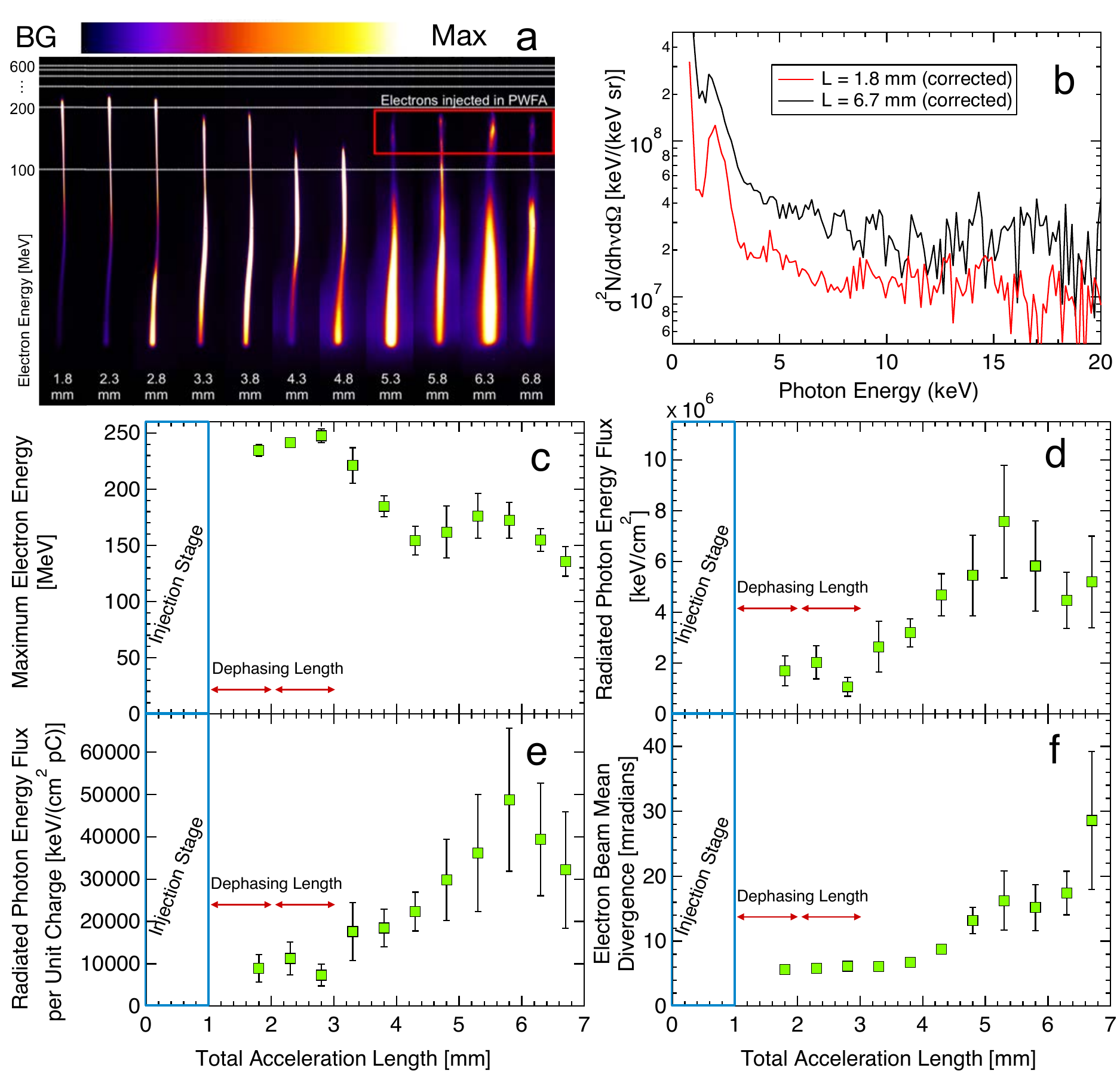}
\caption{\label{fig:exp} (a) Typical electron $p_\parallel-p_\perp$ momentum distributions for various length gas cells measured using a magnetic spectrometer. (b) X-ray spectra at two selected acceleration lengths with quantum efficiency and filter corrections applied. (c) Maximum electron energy, (d) Total radiated photon energy (TRPE), (e) TRPE per unit charge and (f) Electron beam mean divergence. (d) and (e) show a dramatic increase as a function of acceleration length at fixed density. Error bars denote the standard error of the mean.}
\end{figure}

The energy of the electron beam was measured using a 0.8 T magnetic dipole which dispersed the beam onto a LANEX scintillator screen. 
X-ray flux was measured using an Andor iKon-M BR-DD camera, with X-ray sensitivity up to 30 keV and a $13\times13$~mm$^2$ chip size with 1M pixels, that was placed 2.5 m downstream from the interaction region and shielded from the laser light with a 20 $\mu$m Be window. The single hit spectrum was measured using standard single photon counting techniques \cite{Fourmaux_NJP_2011} and adjusted for the quantum efficiency (QE) of the camera and any filters the x-ray beam passes through \cite{Behm2016}, as in Fig.~\ref{fig:exp}(b). The notch in the `corrected' spectra is not a real feature but rather an artifact of uncertainty in the corrections for filter absorption and QE \cite{Behm2016}.

Our main experimental result is a measurement of the photon energy flux as a function of the gas cell acceleration length.  The average laser power was 46 TW, giving an estimated dephasing and depletion length of 1 and 1.9 mm, respectively \cite{Lu2007}.  As shown in Fig. \ref{fig:exp}(a), quasimonoenergetic electron bunches are generated that subsequently dephase and broaden in energy spread. When the acceleration length exceeds \emph{two} dephasing lengths \footnote{The dephasing length is the distance for the electron beam to advance through \emph{half} the accelerating structure.}, the laser energy is depleted for our conditions and the electron beam propagates past the laser front, meaning that it transitions to a beam driven wakefield accelerator \cite{Masson2015}, as evidenced by a second population of electrons being accelerated. In the angular direction, the divergence of the electron bunch is observed to increase considerably after this point (Fig. \ref{fig:exp}(f)). Fig. \ref{fig:exp}(c) shows the electron peak energy from this data averaged over approximately $10$ shots at each length. 

The photon energy flux was simultaneously measured by integrating the signal on the X-ray CCD camera. As can be seen in Fig. \ref{fig:exp}(d), greater than a factor of 3 increase in the photon energy flux occurs on average as the accelerator transitions to beam driven acceleration, albeit with a  constant flux to within error at the longest lengths, suggesting a saturation of the effect. Fig. \ref{fig:exp}(e) shows the X-ray flux per unit charge, \emph{for electrons with energy $>$ 100 MeV only}, indicating this trend is not simply due to an increase in trapped charge. Although the overall beam charge increases with length as well, the bulk of the electrons are decelerated beyond two dephasing lengths and so the high energy component, which contributes predominantly, decreases in charge. The X-ray spectrum at two selected acceleration lengths, Fig. \ref{fig:exp}(b), is also enhanced by more than a factor of 3 on average, consistent with those trends shown in Figs. \ref{fig:exp}(d)-(e). 

This dramatic increase in X-ray flux can be explained by the onset of the electron beam hosing instability. This leads to large amplitude oscillations of the electron beam as indicated by the beam mean divergence based on the average integrated width of the electron beam as a function of propagation length (Fig. \ref{fig:exp}(f)). Since the power radiated by a charge performing large amplitude betatron oscillations scales as $P \simeq r_em_e c^3\gamma^4 k_\beta^4r_\beta^2/3$ \cite{Esarey2002}, an increase in the amplitude of the oscillations $r_\beta$ will lead to a considerable increase in the radiated X-ray power. We can attribute this increase in power to the hosing instability rather than resonant interaction with the laser fields \cite{Thomas2009, Cipiccia_NP_2011} because it occurs at a long propagation distance past a few dephasing lengths. 

Ref. \cite{Whittum1991} describes how coupled motion between an electron beam that is off-center in a pre-formed ion channel and the surrounding plasma sheath electrons may lead to an amplification of an initial transverse offset $x_0$ by a factor ${x(z,\xi)}/{x_0(\xi)} \simeq {0.34}{A^{-3/2}}e^{A}\cos\left[k_\beta z -A/\sqrt{3}+\pi/4\right]
$, 
where the growth factor, $A(z,\xi)$, depends on both the axial displacement $z$, and the ``beam coordinate'' $\xi = t - z/c$, as \cite{Whittum1991,Geraci2000,Huang2007}
$
A(z,\xi) = 1.3[c_rc_{\psi}(k_{\beta}z)^2(\omega_0\xi)]^{1/3} 
$
where $k_{\beta}=k_p/\sqrt{2\gamma}$ ($k_p$ is plasma wavenumber and $\gamma$ is the Lorentz factor), $\omega_0 = k_p/\sqrt{2}$. $c_r$ is a reduction factor for the case of non-adiabatic ion-channel formation, varying along the beam and $c_{\psi}$ is a reduction factor for the case when the plasma electron velocity is large and its magnetic field becomes significant \cite{Huang2007}. 

The growth length $L_g = z/A^{3/2} \approx  0.1\lambda_\beta/\sqrt{c_rc_{\psi}\omega_0\xi}$. For our parameters, $L_g\approx 0.3\lambda_\beta$ for the rear of the beam and $\lambda_\beta\approx300\;{\rm \mu m}$, meaning the expected growth length for the instability is only $\sim 100\;{\rm \mu m}$ and therefore an order of magnitude growth of the offset is expected after a mm propagation. The instability will saturate once the hosing amplitude reaches of order the channel radius. This is broadly consistent with the experimental measurements if the increase in X-ray flux is due to hosing commencing after twice the dephasing limit assuming that this is measured from the end of the injection stage. Recent theoretical studies have shown that the hosing instability may be damped by phase mixing for broad energy spread beams \cite{Mehrling_PRL_2017}, however, if the growth-rate is very large relative to the betatron frequency and the seed amplitude is large, the hosing instability is still significant.



\textsc{Osiris} PIC simulations were conducted to examine the electron acceleration process and to better understand the X-ray radiation enhancement. The simulations were conducted with a preionized gas (n = 0.005 n$_{crit}$, where n$_{crit}$ = $\epsilon m_e \omega_0^2/e^2$ is the critical plasma density). The simulation (82.8 $\times$ 82.8 $\times$ 82.8 $\mu$m$^3$) box was set up in the co-moving frame of the laser, which propagates in the $+z$-direction and is polarized in the $y$-direction. The longitudinal and transverse grid resolution is 31 and 4 cells per laser wavelength, respectively. Absorbing boundary conditions were utilized to prevent electrons reflecting at boundaries. The simulated laser pulse had a 28 fs FWHM pulse duration and focused intensity of 3.4 $\times$ 10$^{19}$ Wcm$^{-2}$ with waist size $w_0 \approx 10 \mu m$. 

In order to perform in-situ radiation simulations, we implemented a Monte Carlo algorithm based method for calculating the classical synchrotron spectrum of the charged particles. This algorithm works by calculating the standard textbook  two-dimensional probability distribution for charged particles in instantaneously relativistic motion \cite{Jacksonbook} to generate macrophotons with randomly determined normalized frequency/normalized angle pair ($\tilde{\omega}, \tilde{\theta}$) for each radiating particle at a particular time step. Whether a particle radiates or not is determined by a minimum Lorentz factor and minimum critical frequency $\omega_{crit}$ based on their trajectory over a timestep. We set $\hbar \omega_{crit} = $ 1.56 keV as the critical frequency of allowed photons ($\omega_{crit}=3\gamma^3 c/\varrho$, where $\varrho$ is the radius of curvature). The emitted macrophotons have the energy that the particle would have radiated over the time step. In the limit of sufficiently large numbers of photons, the algorithm recovers the correct synchrotron spectrum.

\begin{figure}[!ht]
\centering
\includegraphics[width=0.5\textwidth]{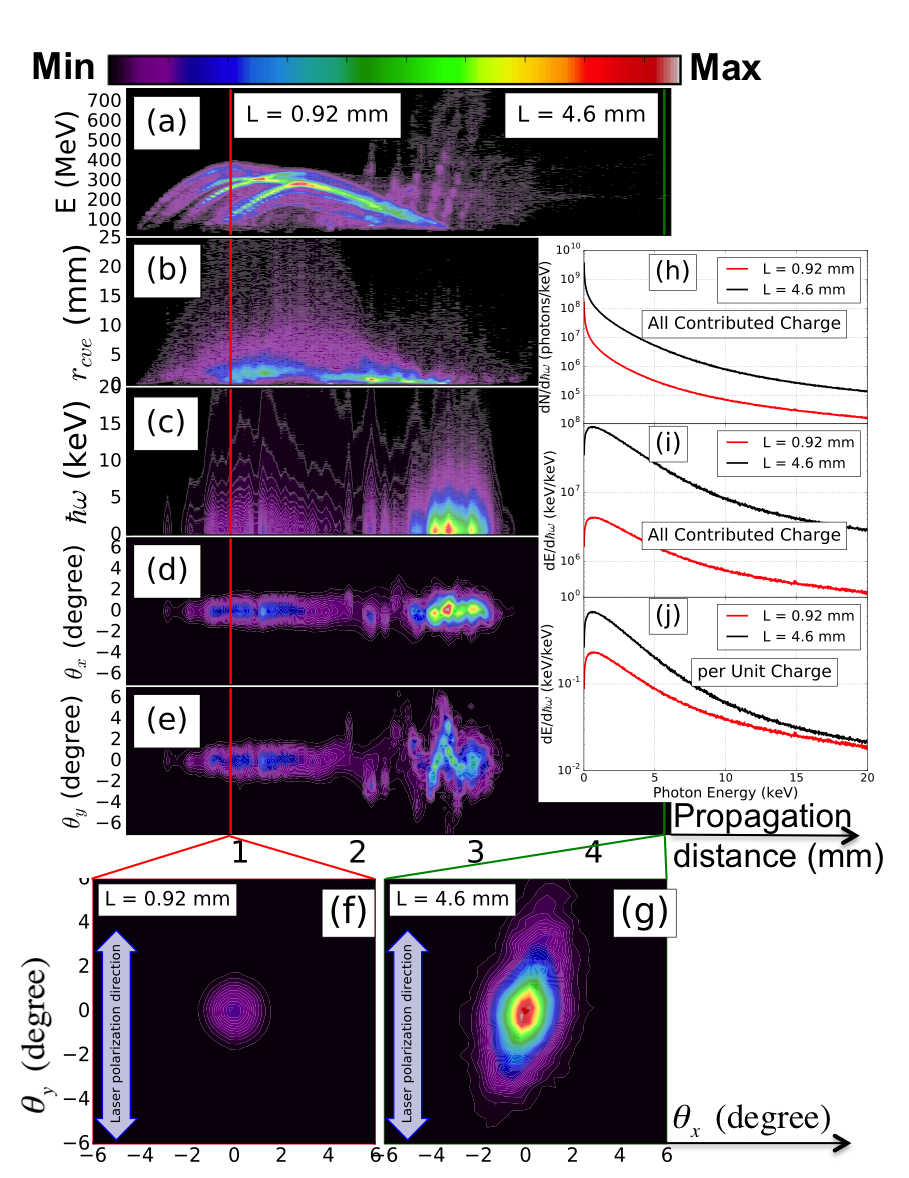}
\caption{\label{fig:merge} Contour plots of phase space density  as a function of  propagation distance $z$ of  (a) electron energy (only those that contribute to the radiation emission), (b) electron radius of curvature ($r_{cve}$) at point of emission, (c) photon energy (keV), (d) photon emission angle $\theta_x$ integrated over $\theta_y$, and (e) photon emission angle $\theta_y$ integrated over $\theta_x$. (f) and (g) show the time integrated X-ray angular distribution at  0.92 mm and 4.6 mm, respectively. The time integrated X-ray spectra of all contributed charge [(h) dN/d$\hbar\omega$ and (i) dE/d$\hbar\omega$ vs. $\hbar\omega$, respectively] and per unit charge [(j) dE/d$\hbar\omega$ vs. $\hbar\omega$] at 0.92 mm and 4.6 mm, respectively.}
\end{figure}

Figs. \ref{fig:merge}(a)-(e) show the time-dependent evolution of electron energy, electron radius of curvature ($r_{cve}$), photon frequency (energy), photon emission angle $\theta_x$ integrated over $\theta_y$, and photon emission angle $\theta_y$ integrated over $\theta_x$. In Fig. \ref{fig:merge}(a), a large group of electrons is initially injected and accelerated in the LWFA regime to energies of several hundred MeV. After propagating past the dephasing length (1.3 mm), the electrons start to decelerate. When the electron beam catches up with the laser pulse at L = 2.8 mm, the net energy gain of the electrons causes pump depletion and the ponderomotive force of the laser is unable to sustain the wakefield. The wakefield becomes beam driven and is capable of trapping and accelerating a secondary bunch of electrons to almost GeV energies. The relatively low density of high energy electrons can be partially caused by the absorbing boundary conditions, which allows electrons to escape.

In Fig. \ref{fig:merge}(b), the electron radius of curvature is observed to decrease with propagation distance. Initially, the radius of curvature is large and the electrons oscillate with a small amplitude in the undulator regime. As the electrons interact with the laser pulse and the electron hosing instability sets in, the electron oscillations increase with a corresponding decrease in the radius of curvature, signifying radiation emission in the wiggler regime \cite{Corde2013}. 

In Figs. \ref{fig:merge}(c)- \ref{fig:merge}(e), the photon energy and the integrated angular distributions are plotted as a function of the propagation distance. The photon energy initially increases and decreases in lockstep with the acceleration and deceleration of the electron beam in the LWFA regime. Afterwards, a significant enhancement occurs for the photon energy due to the acceleration of the secondary electron bunch and its interaction with the laser pulse, which provides a seed perturbation for the electron hosing instability. Figs. \ref{fig:merge}(a) and \ref{fig:merge}(c) are consistent with the experimental data shown in Fig. \ref{fig:exp}. Figs. \ref{fig:merge}(d) and \ref{fig:merge}(e) show that the angular distributions of the photons are relatively isotropic in the earlier stages of the wakefield evolution and signifies electrons oscillating in the undulator regime. As the simulation progresses, the distribution becomes anisotropic, especially in the $\theta_y$ (i.e., laser polarization) direction. The angular deflection in $\theta_y$ reaches approximately $\pm$ 6$^{\circ}$, which is almost three times larger than the deflection in $\theta_x$. The increased angular deflection in the $\theta_y$ direction indicates the onset of the electron hosing instability. The corresponding increase in the photon energy and decrease in the electron radius of curvature can be seen in Figs. \ref{fig:merge}(c) and \ref{fig:merge}(b), respectively.

The integrated angular distributions of the radiated photons are shown in Figs. \ref{fig:merge}(f)-(g). The angular profile of the photons evolves from an isotropic distribution (bottom left) to one that is anisotropic in the laser polarization direction (bottom right). Both the anisotropy and high intensity of the photon angular distribution are caused by the electron hosing instability, which causes the electron beam to emit more radiation at each point in its oscillating trajectory. The integrated X-ray spectra are shown in Figs. \ref{fig:merge}(h)-(j), and mirror the significant increase in both the number of photons and the radiated photon energy (per unit charge) after the instability sets in, consistent with the trend shown in Fig. \ref{fig:exp}(b).

\begin{figure}[!ht]
\centering
\includegraphics[width=0.5\textwidth]{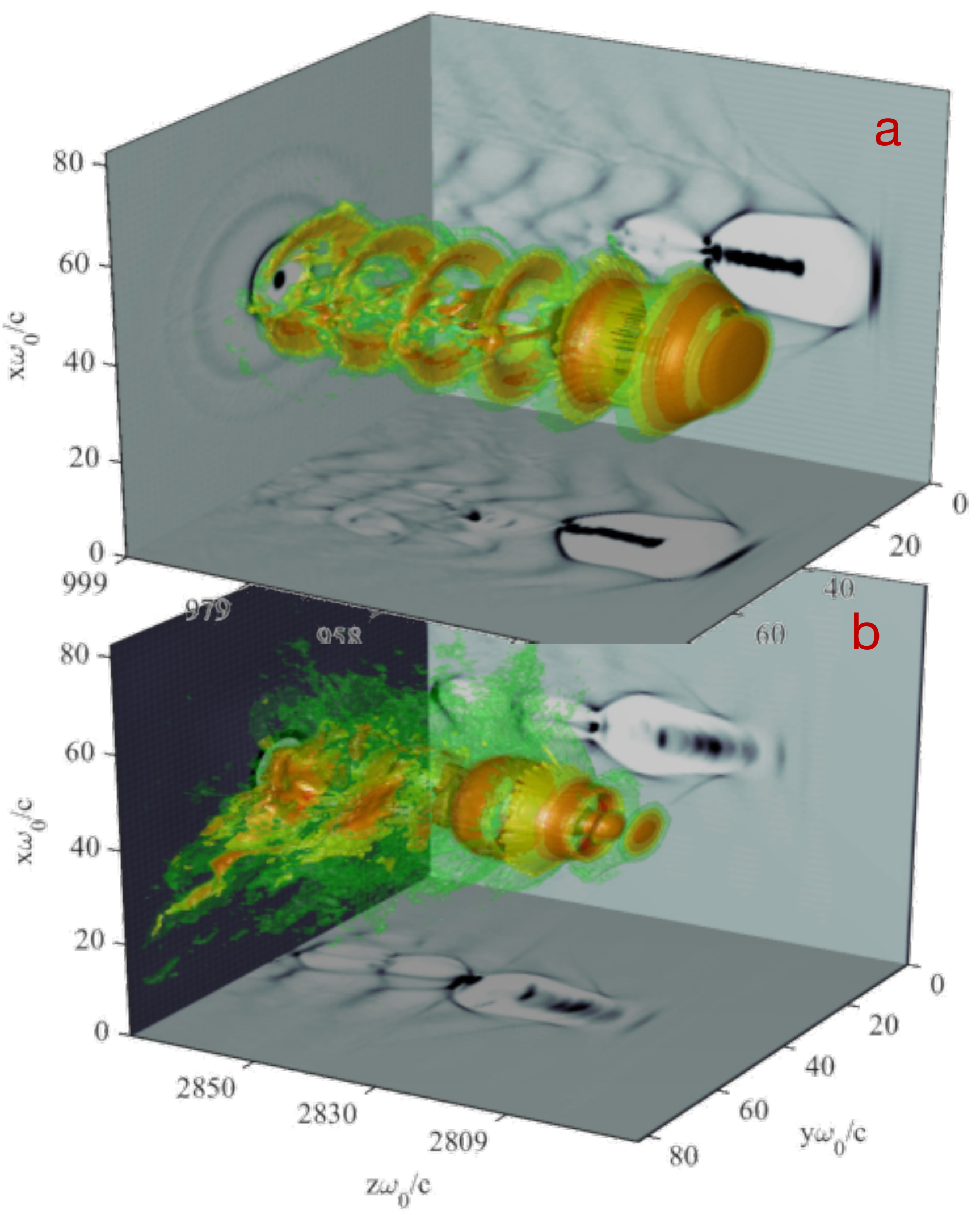}
\caption{\label{fig:3d} 3D map of charge-density and its projection on the $x-z$ and $y-z$ planes. At (a) $z=$0.92 mm, (b) $z=$2.8 mm.}
\end{figure}

A 3D charge-density map showing the evolution of the wakefield at different lengths is given in Fig. \ref{fig:3d}. In Fig. \ref{fig:3d}(a), the wakefield is in the LWFA regime and electrons are continuously injected into the first bubble. A clear electron beam is formed as seen in the projected images. As this first group of electrons catches up and interacts with the laser pulse front, Fig. \ref{fig:3d}(b), the electrons undergo increased oscillations in the laser polarization direction. This motion disrupts the spherical shape of the cavity, especially towards the cavity rear. In the $y-z$ plane, a noticeable deviation of the electron beam from the cavity axis can be seen. This slight deviation plants the seed for the electron hosing instability. A secondary group of electrons also forms at the rear of the cavity in Fig. \ref{fig:3d}(b), which is then accelerated by the primary bunch after a transition from an LWFA to a PWFA occurs \cite{Masson2015}.

In conclusion, an enhancement in the X-ray emission resulting from the electron hosing instability has been demonstrated. Both experimental and 3D PIC simulation results show that the total radiated photon energy starts to increase shortly after the electron beam catches up with the laser pulse. The interaction of the primary electron beam with the laser pulse provides a seed perturbation for the electron hosing instability and thus significantly enhances the X-ray radiation by almost an order of magnitude. Subsequently, the wakefield driven by the high-energy electron beam can trap and accelerate a secondary group of electrons, during which a transition from an LWFA to a PWFA occurs. Furthermore, this enhancement cannot be attributed to simply an increase in the total number of accelerated charges with length. The onset of the instability can also be inferred from the simulations by observing the evolution of the emitted photon angular distribution and the electron radius of curvature. The photon angular distribution changes from an approximately isotropic distribution to an anisotropic distribution, where the emitted angle $\theta_y$ (in the laser polarization direction) is about three times larger than $\theta_x$, with a corresponding increase in the X-ray source size. At the same time, the electron radius of curvature shows a negative linear correlation with the variation of $\theta_y$ due to the transition from the undulator regime to the wiggler regime. The greatly enhanced X-ray radiation can be achieved by taking advantage of a naturally occurring instability in wakefield acceleration and is useful for applications where a high X-ray flux is required with minimal change to experimental design.

This work is supported by the U.S. Department of Energy/National Nuclear Security Administration grant DE-NA0002372, the National Science Foundation Career grant 1054164, and the Air Force Office of Scientific Research Young Investigator Program grant FA9550-12-1-0310. The author would like to acknowledge the \textsc{Osiris} consortium for use of the \textsc{Osiris} 2.0 framework.


\end{document}